\documentclass[twocolumn]{aastex62}

\usepackage{epsfig}
\usepackage{amsmath}
\usepackage{enumerate}
\usepackage{mathrsfs}
\usepackage{natbib}
\usepackage{units}
\usepackage{hyperref}
\usepackage{color}
\definecolor{orange}{RGB}{220,110,0}
\definecolor{violet}{RGB}{250,100,250}

\def\mis{\xi}
\def\Rlc{R_{\rm lc}}
\def\Lsd{L_{\rm sd}}
\def\tc{t_{\rm s,cool}}

\graphicspath{{./}{figures/}}

\received{June 1, 2019}
\revised{June 7, 2019}
\accepted{\today}
\submitjournal{ApJ}

\shorttitle{Pulsar wind-heated disk and modes in PSR~J1023+0038}
\shortauthors{A. Veledina, J. N\"{a}ttil\"{a} \& A. M. Beloborodov}

\begin{document}

\title{Pulsar wind-heated accretion disk and the origin of modes in transitional millisecond pulsar PSR~J1023+0038}

\author[0000-0002-5767-7253]{Alexandra~Veledina}
\affiliation{Department of Physics and Astronomy, FI-20014 University of Turku, Finland}
\affiliation{Nordita, KTH Royal Institute of Technology and Stockholm University, Roslagstullsbacken 23, SE-10691 Stockholm, Sweden}
\affiliation{Space Research Institute of the Russian Academy of Sciences, Profsoyuznaya Str. 84/32, Moscow, 117997, Russia}

\author[0000-0002-3226-4575]{Joonas N\"{a}ttil\"{a}}
\affiliation{Nordita, KTH Royal Institute of Technology and Stockholm University, Roslagstullsbacken 23, SE-10691 Stockholm, Sweden}

\author{Andrei M. Beloborodov}
\affiliation{Physics Department and Columbia Astrophysics Laboratory, Columbia University, 538 West 120th Street, New York, NY 10027, USA}
\affiliation{Max Planck Institute for Astrophysics, Karl-Schwarzschild-Str. 1, D-85741, Garching, Germany}

\begin{abstract}
Transitional millisecond pulsars provide a unique set of observational data for understanding accretion at low rates onto magnetized neutron stars. 
In particular, PSR~J1023+0038 exhibits a remarkable bimodality of the X-ray luminosity (low and high modes), pulsations extending from the X-ray to the optical band, GeV emission, and occasional X-ray flares.
We discuss a scenario for the pulsar interaction with the accretion disk capable of explaining the observed behavior.
We suggest that during the high mode the disk is truncated outside the light cylinder, allowing the pulsar wind to develop near the equatorial plane and strike the disk. 
The dissipative wind-disk collision energizes the disk particles and generates synchrotron emission, which peaks in the X-ray band and extends down to the optical band.
The emission is modulated by the pulsar wind rotation, resulting in a pulse profile with two peaks 180\degr\ apart. 
This picture explains the high-mode luminosity, spectrum, and pulse profile (X-ray and optical) of PSR~J1023+0038. 
It may also explain the X-ray flares as events of sudden increase in the effective disk cross section intercepting the wind.
In contrast to previously proposed models, we suggest that the disk penetrates the light cylinder only during the low X-ray mode. 
This penetration suppresses the dissipation caused by the pulsar wind-disk collision, and the system enters the propeller regime.
The small duty cycle of the propeller explains the low spindown rate of the pulsar.
\end{abstract}


\section{Introduction} \label{sec:intro}

Transitional millisecond pulsars constitute a subclass of neutron stars which swing between the radio-pulsar and accretion states.
They provide strong evidence for the recycling scenario for the rotation-powered millisecond pulsars, where the occasional accretion episodes lead to the neutron star spin-up \citep{BKK76,Alpar82}. 
To date, three sources have been seen to switch between the states: PSR~J1023+0038 \citep{Archibald09}, IGR~J18245--2452 \citep{Papitto13} and XSS~J12270--4859 \citep{Bassa14}.
In particular, the wealth of multiwavelength campaigns for PSR~J1023+0038 revealed rich variety of observed phenomena and allowed its detailed study.
We focus on this object in this paper.

Optical observations in the early 2000s identified the previously known radio source as a cataclysmic variable \citep{Bond02,Szkody03}, as the optical spectra demonstrated double-peak emission lines, the hallmark of accretion disks.
In 2003, the emission lines disappeared and the source was later discovered as a radio pulsar \citep{ThorArm05,Archibald09}.
For a decade, PSR~J1023+0038 stayed in the radio pulsar (RP) state, until 2013, when the radio pulsations disappeared and the optical/UV, X-ray and $\gamma$-ray activity enhanced, suggesting reactivation of accretion \citep{Stappers14,Patruno14}.
The system entered the low-mass X-ray binary (LMXB) state and has remained in it since then.

In the LMXB state, PSR~J1023+0038 exhibits two distinct X-ray modes \citep{Bogdanov15,Archibald15}: the high mode with luminosity $L \sim 3\times10^{33}$~erg~s$^{-1}$ (present $\sim$80\% of time), and the low mode with $L \sim 5\times10^{32}$~erg~s$^{-1}$ (about $20\%$ of  time). 
In addition, X-ray flares have been detected with luminosity up to $L\sim 10^{34}$~erg~s$^{-1}$ ($<2\%$ of time).
X-ray spectra in all the modes are well described by a power law with nearly constant photon index: $\Gamma\approx1.7$ in the high and flaring modes, and $\Gamma\approx 1.8$ in the low mode \citep{Bogdanov15}.

Two distinct modes are also identified in the UV band \citep{Jaodand16,CotiZelati18}.
Optical and infrared emission also shows two modes similar to the X-ray modes \citep{Shahbaz15,Papitto19},  at least for some time \citep{Kennedy18,Shahbaz18}.
Furthermore, optical polarization and spectroscopic characteristics change between the modes \citep{Hakala18}, in addition to the long-term optical variability at the system orbital period \citep{CotiZelati14,Papitto18}.
Interestingly, coherent pulsations at the pulsar spin frequency were detected in both X-rays and optical wavelengths in the high mode, but not in the low mode \citep{Archibald15,Bogdanov15,Ambrosino17,Papitto19}.
Radio activity was found to anti-correlate with the X-ray modes. 
Radio flares are typically detected during the X-ray low mode,  although they were sometimes seen also in the high and flaring modes \citep{Bogdanov18}.

A number of models for this rich behavior have been discussed.
The increase of the X-ray and $\gamma$-ray emission after the switch to the LMXB state in 2013 was first associated with the enhanced wind from the donor star and its collision with the pulsar wind -- the intra-binary shock \citep{Takata14,Li14,CotiZelati14}.
The discovery of the bimodal flux distribution \citep{Patruno14,Bogdanov15} triggered suggestions of accretion in the propeller regime \citep{PapittoTorres15,Campana16}, with significant accretion onto the magnetic poles of the neutron star during the high mode and reduced accretion during the low mode.

A major challenge faced by the accretion picture is the observed spindown rate of the neutron star.
Accretion is expected to strongly change the pulsar spindown rate from its value in the RP state.
By contrast, only $\sim$27\% increase of the spindown rate was found during the LMXB state \citep{Jaodand16}.
Another puzzle is the origin of optical pulsations \citep{Ambrosino17}.

In this paper, we propose a new scenario capable of explaining the multiwavelength behavior of the three modes in PSR~J1023+0038.
We suggest that accretion onto the neutron star is negligible in the high mode. 
Instead, the high-mode luminosity is sustained by the pulsar wind collision with the accretion disk, which is kept outside the light cylinder. 
Thus, the high mode is {\it powered by the neutron star rotation} rather than by accretion.
Occasional enhancements of the cross-section for the wind-disk interaction increase the dissipated power and can lead to X-ray flares. 
We argue that in the low mode the disk penetrates the light cylinder, and this reduces its dissipative interaction with the pulsar wind. 
Episodes of significant accretion onto the magnetic poles might occur during the low mode, however most of its power is generated far from the star by accretion in the propeller regime.

We outline the proposed picture for the low/high and flaring modes in Section~\ref{sect:pheno}, and then discuss in more detail the radiative mechanism of the high mode, its luminosity, spectrum and pulse profile in Section~\ref{sect:phys}.
Our scenario is further discussed and compared to observations in Section~\ref{sect:discus} and summarized in Section~\ref{sect:summary}.

\section{Motivation and outline of the proposed picture} \label{sect:pheno}

PSR~J1023+0038 is a binary system with orbital period $P_{\rm orb}=4.75$~hr and a donor star of mass $M_{2}\sim0.24 M_{\odot}$ \citep{McConnell15}.
Radio data gave measurements for its distance $d=1.37\pm0.04$~kpc and the neutron star mass estimate $M_{\rm NS}=1.71\pm 0.16 M_{\odot}$ \citep{Deller12}.

With spin period $P_{\rm spin}=1.69$~ms \citep{Archibald09}, PSR~J1023+0038 is one of the fastest known pulsars. 
The spindown rate observed during the RP state $\dot{P}_{\rm RP}=(5.30\pm0.07)\times10^{-21}$ \citep{Deller12} implies that the neutron star has the magnetic dipole $\mu\approx 10^{26}$~G~cm$^3$. 
The pulsar light cylinder radius, 
\begin{equation}
 \Rlc = \frac{c P_{\rm spin}}{2\pi} \approx 80 {\rm~km},
\end{equation}
exceeds the corotation radius by a factor of 3,
\begin{equation}
  R_{\rm co} = \left( \frac{GM_{\rm NS} P_{\rm spin}^2}{4 \pi^2} \right)^{1/3} \approx 24 {\rm ~km}.
\end{equation}
The Alfv\'{e}n radius 
\begin{equation}
  R_{\rm A}= \mu^{4/7} ( 2 G M_{\rm NS} \dot{M}^2 )^{-1/7}
\end{equation}
depends on the accretion rate $\dot{M}$, which is usually related to the observed luminosity and radiative efficiency of accretion.
If accretion proceeded to the neutron star surface, it would generate luminosity $L_{\rm X}\sim GM_{\rm NS}\dot{M}/R_{\rm NS}$, and one would find $R_{\rm A}\approx 100\, (L_{\rm X}/10^{34}\,{\rm erg~s^{-1}})^{-2/7}$~km. 
This does not give a consistent accretion picture for PSR~J1023+0038, as it implies $R_{\rm A}>\Rlc$.
For the high- and low-mode $L_{\rm X}$ one would find $R_{\rm A}/\Rlc\sim 2-3$, and so the disk is not expected to enter the light cylinder. 

\begin{figure*}
\plotone{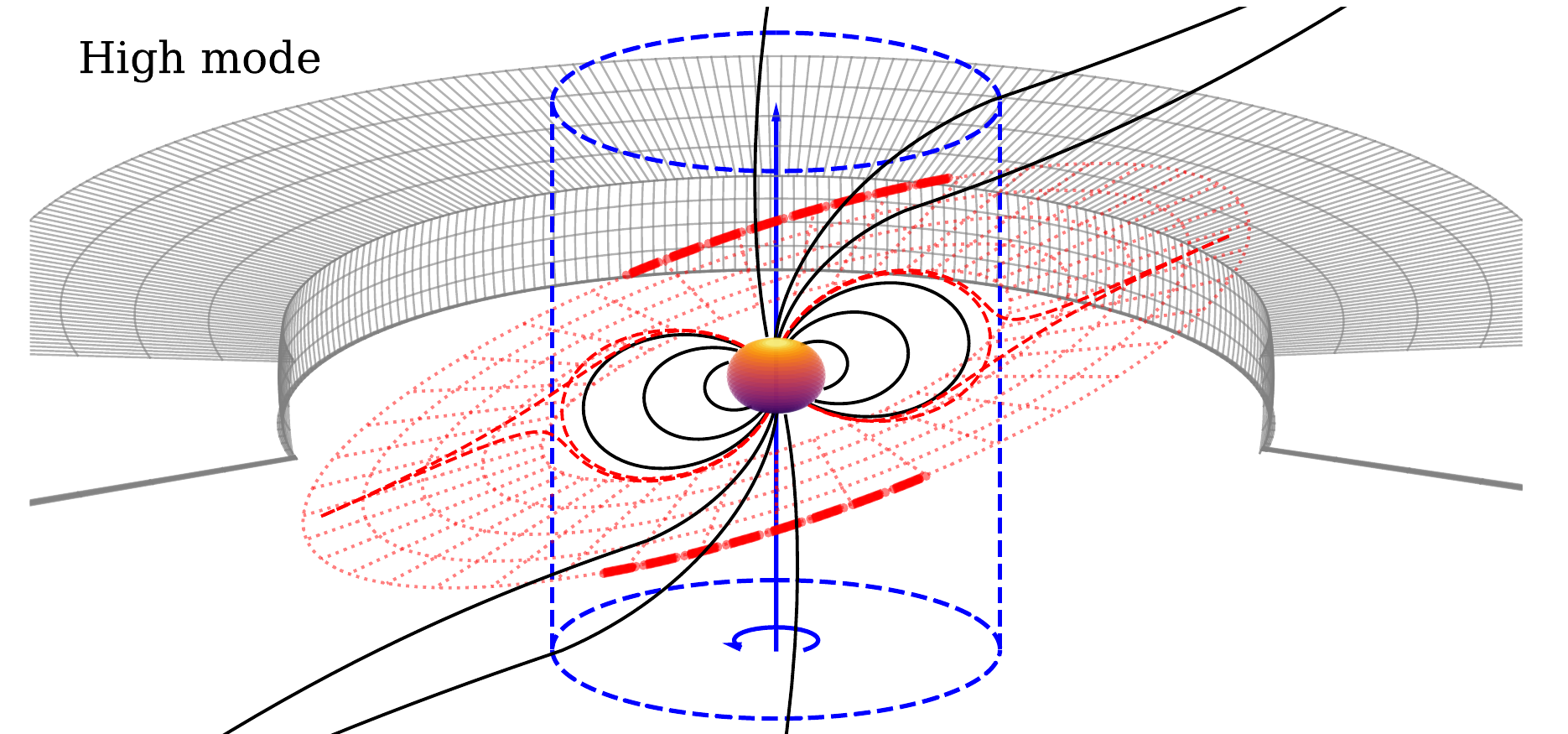}
\caption{
Schematic picture of the high mode. 
The disk stays outside the light cylinder (dashed blue) and interacts with the pulsar wind, in particular with its wobbling equatorial current sheet (red dotted). 
Energy dissipation in the wind-disk collision peaks at the intersection of the current sheet and the inner boundary of the disk (thick red dashed). 
The two dissipation peaks result in double-peak pulsations as the pulsar rotates (see text).
\label{fig:geom_high}
}
\end{figure*} 

\begin{figure*}
\plotone{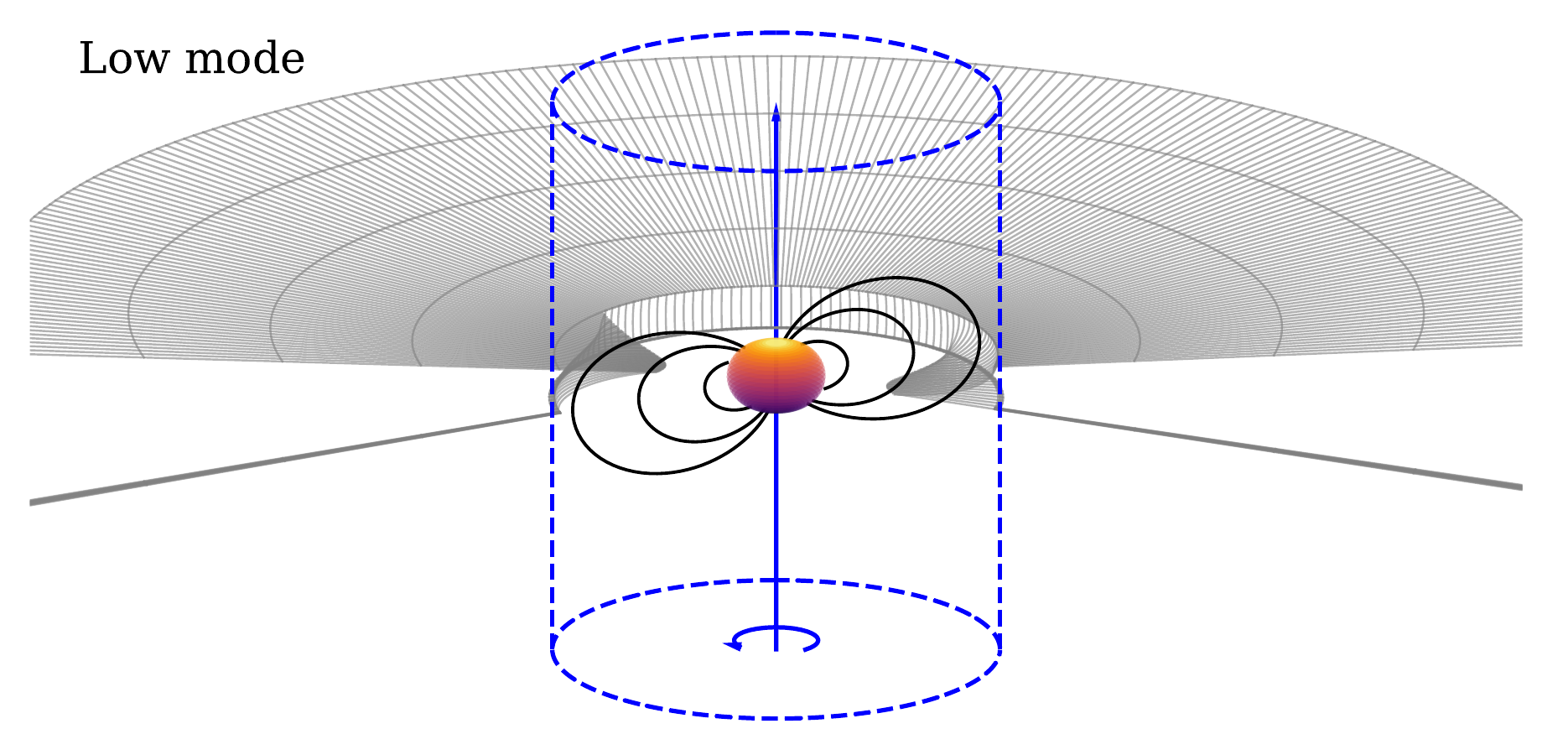}
\caption{
Schematic picture of the low mode.
The disk penetrates the light cylinder (dashed blue) and interacts with the closed magnetosphere (solid black).
\label{fig:geom_low}
}
\end{figure*}

When $R_{\rm A}\gtrsim \Rlc$, the pulsar wind is powerful enough to keep the accretion disk outside the light cylinder most of the time, with occasional penetrations of the disk matter inside $\Rlc$, as demonstrated by recent numerical simulations \citep{PT17}.
The opposite regime $R_{\rm A}<\Rlc$ is possible in the high (and low) modes of PSR~J1023+0038 only if the observed luminosity is produced by a much higher accretion rate $\dot{M}$ with a low radiative efficiency.
This assumption might be self-consistent if the accreting gas does not reach the neutron star surface, and instead is ejected in the propeller regime \citep{IS75}.

To explain the observed high/low mode dichotomy, several scenarios have been proposed.
Some models invoke the propeller regime of accretion \citep{PapittoTorres15,Campana16}, and others invoke a fast switch between magnetospheric accretion (high mode) and the intra-binary shock emission (low mode) \citep{Linares14,CotiZelati18}.
The X-ray pulsations observed in the high mode were usually associated with partial capturing of matter by the neutron star magnetosphere and channeled accretion onto the magnetic poles. 

The main questionable assumption of the propeller scenario for PSR~J1023+0038 is that the accretion disk is truncated near the corotation radius ($R_{\rm A}\sim R_{\rm co}$).
This is required to avoid a strong torque on the star and make it consistent with spindown observations.
The condition $R_{\rm A}\sim R_{\rm co}$ corresponds to accretion rate $\dot{M}\sim 10^{-3}\dot{M}_{\rm Edd}$ (expressed in the Eddington units), which is typical for accreting millisecond pulsars. 
At this rate, blackbody disk emission would be clearly seen in the X-ray band \citep{PRA09,KIA11}, in contrast to the observed spectra in PSR~J1023+0038.
If accretion to $R_{\rm co}$ is assumed to proceed in a radiatively inefficient regime, $\dot{M}\sim 10^{-3}\dot{M}_{\rm Edd}$ would still imply a minimum X-ray luminosity \citep{YN14}, which would exceed the observed high-mode luminosity $L_{\rm X}\sim 3\times 10^{33}$~erg~s$^{-1}$.

Furthermore, any scenario explaining X-ray pulses by accretion onto the neutron star magnetic poles faces the following problem.
The pulse profiles during the LMXB state are approximately sinusoidal at a double pulsar spin frequency, quite different from the pulses observed during the RP state \cite[see figure~2 in][]{Archibald15}.
To reproduce the double peaks with close amplitudes, the two poles heated by accretion must be located at low latitudes and the pulsar spin axis must have a high inclination of $\sim$80\degr\ with respect to the line of sight \citep{VP04,NP18}.
One problem for such high inclinations is that in the presence of an accretion disk the secondary pulse becomes blocked \citep{IP09}.
Yet more serious problem is that a high inclination would be in conflict with the measured value $i=42\pm2$\degr \citep{Archibald13}. 

\begin{figure}
\plotone{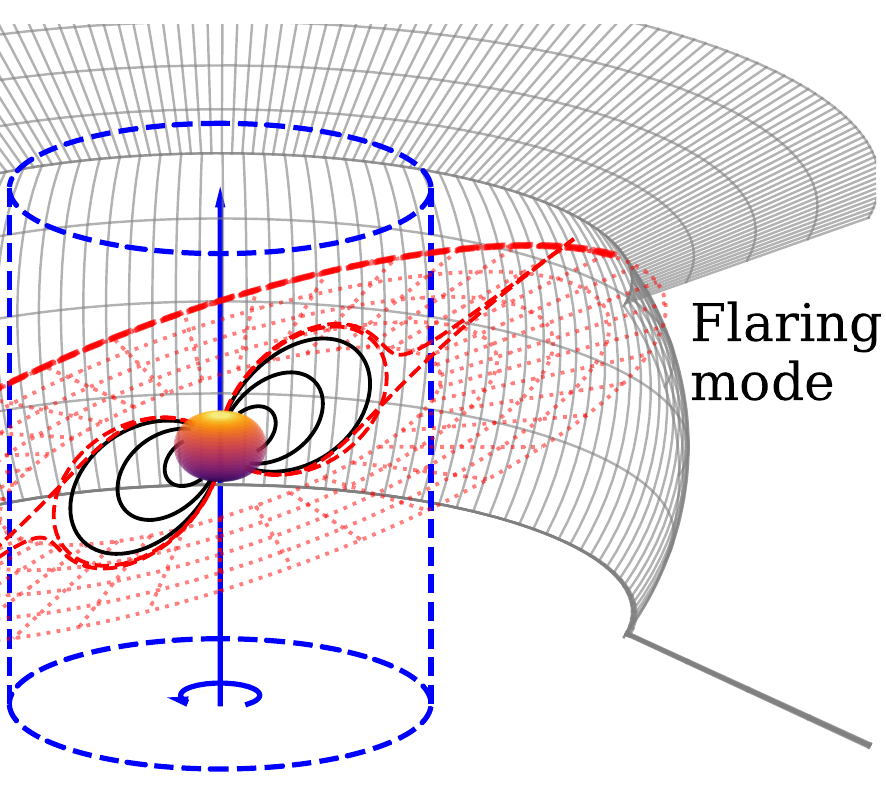}
\caption{
Schematic picture of the flaring mode. 
The disk becomes geometrically thick and now intercepts most of the pulsar wind power, resulting in a higher luminosity. 
Pulsation amplitude is reduced.
\label{fig:geom_flare}
}
\end{figure}

This motivates us to consider an alternative picture.
We suggest that the accretion rate remains rather low, $\dot{M}\sim 10^{-5} M_{\rm Edd}$, in both high and low modes, so that the disk truncation radius always stays around $\Rlc$ and well
outside $R_{\rm co}$.
The system exhibits the high mode when the disk inner edge is outside $\Rlc$ (Figure~\ref{fig:geom_high}).
The time-averaged spindown torque remains small, comparable to that in the RP state, because the system spends 80\% of the time in the high mode, when the pulsar spindown is not affected by the disk.

The disk occasionally penetrates the light cylinder, and this penetration causes the transition to the low mode (Figure~\ref{fig:geom_low}).
In the penetration regime, almost all matter supplied by the disk is still ejected\footnote{There may be episodes during the low mode when a stream of matter falls to the magnetic poles of the neutron star.
However, such episodes must be rare, as no oscillations at the star rotation frequency were found in the low mode \citep{Archibald15,Bogdanov15}.}, now by the propeller effect of the closed magnetosphere \citep{IS75}.
In this regime, the observed luminosity can be powered by both the spindown power of the pulsar and the accretion power.
 
The high mode in our scenario corresponds to the disk staying outside the light cylinder (see Figure~\ref{fig:geom_high} and \citealt{Papitto19} for a similar picture of the high-mode geometry).
We argue below that in this configuration a strong equatorial pulsar wind bombards the inner edge of the disk, enhancing the dissipation rate and the X-ray luminosity from the system. 

The effective cross section for the dissipative wind-disk collision depends on the scale-height $H$ of the disk inner edge.
Low-$\dot{M}$ accretion can be radiatively inefficient, and such accretion disks normally have a hydrostatic scale-height comparable to radius. 
However, at the inner edge, where the disk attempts to penetrate the pulsar wind, $H$ is shaped mainly by the collision with the wind rather than the hydrostatic balance, and $H/R$ can be significantly below unity most of the time.

As known from the studies of isolated pulsars, the wind power has a strong fan-like peak at/around the current sheet that extends outside the light cylinder \citep[see figure~4 in][]{ChenBeloborodov14}.
When the rotation axis and the magnetic dipole moment of the pulsar are misaligned by some angle $\mis$, the current sheet oscillates with the period $P_{\rm spin}$ around the equatorial (orbital) plane, and occupies latitudes $<\mis$. 
The misalignment implies that the fan of the wind peak slides along the inner edge of the  disk as the pulsar rotates. 
We propose that this effect gives rise to the observed oscillations of the X-ray flux in the high mode with the period $P_{\rm spin}/2$.
The energy dissipated in the wind-disk collision is radiated via synchrotron (and inverse Compton) mechanism, producing an extended power-law spectrum.

We also suggest that the basic picture and the emission mechanism of the flaring mode is similar to the high mode, except that the area of the wind-disk collision is significantly increased, because the disk becomes thick, $H/R\sim 1$ (Figure~\ref{fig:geom_flare}).
Strong changes in the disk thickness exposed to the pulsar wind have been observed in the magnetohydrodynamic (MHD) simulations of \citet{PT17}, in particular in their $\mu=80$ and $\mu=160$ models. 
 
This scenario of the low, high, and flaring modes does not require strong variations in the matter supply to the accretion disk for transitions between the modes. 
Assuming that the Alfv\'en radius $R_{\rm A}$ is at a few $\Rlc$, we obtain an estimate for accretion rate $\dot{M}\sim10^{13}-10^{14}$~g~s$^{-1}$.
A disk truncated at $R\sim \Rlc$ and accreting at such rates is capable of producing luminosities $\sim 10^{32}-10^{33}$~erg~s$^{-1}$, depending on the radiative efficiency. 
Thus, the observed low-mode luminosity can be sustained by both accretion and the spindown power.

The characteristic electron density inside the accretion disk can be estimated as follows,
\begin{equation}
 n_{\rm disk} = \frac{\dot{M}}{ 4\pi \Rlc H v_{\rm acc} m_{\rm p}} \sim \left(10^{15}-10^{16} \right) \left(\frac{H/R}{0.1}\right)^{-3} {\rm cm}^{-3}.
\end{equation}
Here $v_{\rm acc}\sim\alpha (H/R)^2 v_{\rm K}$ is the accretion velocity in a disk of half-thickness $H$ rotating with Keplerian speed $v_{\rm K}=(GM_{\rm NS}/\Rlc)^{1/2}$, $\alpha\sim 0.1$ is a viscosity coefficient \citep{SS73}, and $m_{\rm p}$ is the proton mass.
Thomson optical depth across the disk is then
\begin{equation}
  \tau = 2H\sigma_{\rm T} n_{\rm disk} \sim  \left(10^{-2}-10^{-3} \right) \left(\frac{H/R}{0.1}\right)^{-2} \ll 1, 
\end{equation}
and so the disk is optically thin.

We expect only sporadic accretion onto magnetic poles, as the matter is mostly expelled from the system.
Hence, we do not expect any significant emission from the neutron star surface.
The main source of emission is the optically thin plasma of the truncated accretion disk, which can efficiently radiate through synchrotron self-Compton (SSC) mechanism, in particular when it is energized by the collision with the plasma wind.

\section{Powering the high mode with the pulsar wind}\label{sect:phys}

\subsection{Energetics}\label{sect:ener}

The configuration with the disk outside the light cylinder can be most dissipative and produce X-ray luminosity exceeding the accretion power. 
Therefore, we associate this configuration with the high mode. 
The basic argument can be seen in the idealized case of an aligned rotator.
When the disk is outside $\Rlc$, the equatorial pulsar wind develops near the magnetospheric Y-point at radius $R_{\rm Y}\approx \Rlc$, and then strikes the disk, depositing its power into the disk plasma.
The high mode of PSR~J1023+0038 in this picture is powered by the rotation energy extracted from the pulsar, carried by its wind, and eventually radiated by the disk.

The magnetospheres and winds of isolated pulsars have recently been studied with global kinetic plasma simulations (see \citealt{CerittiBeloborodov17} for a review).
The simulations show that 15--20\% of the wind power is dissipated just outside the Y-point of the aligned rotator and forms energetic plasmoids filled with copious $e^\pm$ pairs \citep{ChenBeloborodov14,PSC15}.
No kinetic plasma simulations have been done yet for pulsars surrounded by an accretion disk; only force-free \citep{Parfrey17} and MHD \citep{PT17} models are available.
The wind dissipation is expected to increase in the presence of a disk outside the light cylinder, as the disk forms an obstacle for the pulsar wind. 
The disk also provides many more particles that can radiate the dissipated power in the X-ray band.
One may expect that the dissipation of the spindown power is reduced when the disk penetrates the light cylinder and occupies the Y-point region, so that it begins to interact with the closed rotating magnetosphere rather than a pair-loaded relativistic wind. 

The pulsar wind power is known from observations during the RP state, when there is no accretion, \citep{Deller12}
\begin{equation}
  L_{\rm sd}\approx 4 \times 10^{46} \dot{P}_{\rm RP}P^{-3} \sim 4.4 \times10^{34}{\rm ~erg~s}^{-1}.
\end{equation}
The corresponding power computed from the average spindown rate $\dot{P}$ during the LMXB state is somewhat higher, $L_{\rm sd, LMXB}\sim7\times10^{34}$~erg~s$^{-1}$ \citep{Jaodand16}.
We interpret the increase of spindown torque as the result of occasional penetration of the accretion disk into the light cylinder, when the open magnetic flux of the pulsar, and hence its wind power, increases \citep{Parfrey16,PT17}.

The angular distribution of the wind power depends on the misalignment angle $\mis$ between the magnetic dipole moment $\boldsymbol{\mu}$ and the pulsar angular velocity $\boldsymbol{\Omega}$. 
The angular distribution at the radius of interest, $R\approx 2R_{\rm lc}$, was calculated in the MHD models of isolated pulsars \citep[see figure~3 in][]{Tchekhovskoy2013} and full kinetic plasma models \citep[see figure~4 in][]{ChenBeloborodov14}. 
For aligned (or anti-aligned) rotators, the wind power has a strong narrow spike at the equator. 
This spike is dominated by the plasma energy rather than the Poynting flux. 
The spike results from the equatorial dissipation of magnetic energy near the Y-point, and carries $\sim$15--20\% of the total wind power. 
Poynting flux from aligned/anti-aligned rotators has a broad, smooth peak at angles $\theta\sim 60\degr$ from the rotation axis ($30\degr$ from the equatorial plane). 

In inclined rotators, the angular distribution of the wind power is modulated with  rotational phase. 
It is still expected to have a strong peak in the dissipative equatorial current sheet, however, now the sheet is wobbling around the equatorial plane. 
When averaged over the rotational phase the power peak becomes less concentrated but remains well above the average value $L_{\rm sd}/4\pi$. 

The solid angle $\Omega_{\rm disk}$ subtended by the ``obstacle'' --- the accretion disk --- depends on its effective
thickness $2H$. 
For instance, $H/R=0.1$ gives $\Omega_{\rm disk}/ 4\pi \sim 0.1$. 
A conservative estimate for the power dissipated in the wind-disk collision is obtained assuming isotropic wind, $L_{\rm diss}\sim (\Omega_{\rm disk}/4\pi)\Lsd$. 
Thus, the simplest estimate for the dissipated fraction of the wind power is
\begin{equation}
    \zeta_{\rm w}\equiv\frac{L_{\rm diss}}{L_{\rm sd}} \gtrsim \frac{H}{R}.
\end{equation}
This gives $L_{\rm diss}\gtrsim 4.4\times10^{33}$~erg~s$^{-1}$, sufficient to generate the high-mode luminosity $L_{\rm X}\sim 3\times 10^{33}$~erg~s$^{-1}$ with the radiative mechanism discussed below.

In a more accurate model, $\zeta_{\rm w}$ depends on the misalignment angle $\mis$. 
For small $\xi$, even a relatively thin disk residing in the equatorial plane can intercept the 15--20\% of the wind power concentrated in the current sheet.
When $\tan \mis>H/R$ (as shown in Figure~\ref{fig:geom_high} for the high mode), the disk does not intercept the entire current sheet.

\subsection{Emission mechanism and spectrum}\label{sect:emis}

The collision between the pulsar wind and the disk involves shocks and likely deposits energy into the disk turbulence. 
It might also involve magnetic reconnection. 
Here we outline a simplified radiative model agnostic to the dissipation mechanism, and discuss emission from the wind and disk particles energized by the collision.

The average energy per particle in a pulsar wind is given by 
\begin{equation}
    \eta=\frac{L_{\rm sd}}{\dot{N}_{\rm w} m_{\rm e} c^2},
\end{equation}
where $L_{\rm sd}\approx \mu^2\Omega^4/c^3$ is the wind power, $\dot{N}_{\rm w}={\cal M} I/e$ is the rate of particle ejection in the wind ($e$ is the electron charge), $I\approx\mu\Omega^2/c$ is the electric current flowing along open magnetic field lines \citep{GJ69}, and ${\cal M}\sim 10-10^3$ is a factor describing the multiplicity of pair creation in the magnetosphere \citep{Sturrock71}.
Dissipation of a large fraction of the intercepted $L_{\rm sd}$ envisioned by our scenario would imply that the wind is heated to a thermal Lorentz factor $\gamma_{\rm m, w}\sim\eta$, which is given by
\begin{equation}\label{eq:gamma_mw}
    \gamma_{\rm m, w} \sim \frac{e\mu\Omega^2}{{\cal M} m_{\rm e} c^4}\sim \frac{10^9}{\cal M}.
\end{equation}
This Lorentz factor is huge for any plausible multiplicity ${\cal M}$. 
Emission from such energetic particles is in the far gamma-ray band and does not give a high radiative efficiency in the X-ray band. 
We conclude that the observed X-ray luminosity cannot be emitted by heated wind particles.

Next consider the disk particles.
The particles inflow in the equatorial plane to the disk termination radius and get ejected as a result of energy deposition by the wind. 
The particle flow rate in this circulation is $\dot{N}\sim \dot{M}/m_{\rm p}\sim 10^{37}\,$s$^{-1}$. 
The average energy deposited by the wind per disk particle is 
\begin{equation}\label{eq:gamma}
    \bar{E}=\frac{\zeta_{\rm w} L_{\rm sd}}{\dot{N}}.
\end{equation}
If a fraction $\zeta_{\rm d}$ of the total flow of disk particles $\dot{N}$ is heated impulsively in the wind-disk interaction region, and a significant fraction $\epsilon_{\rm e}$ of the heat is given to electrons (as would occur in a shock), then the mean electron Lorentz factor immediately after heating is $\gamma_{\rm m}=\epsilon_{\rm e}\bar{E}/\zeta_{\rm d} m_{\rm e} c^2$. 
This gives
\begin{equation}\label{eq:gamma_m}
    \gamma_{\rm m}\sim \epsilon_{\rm e} \frac{m_{\rm p}}{m_{\rm e}}
    \frac{\zeta_{\rm w} L_{\rm sd}} {\zeta_{\rm d} \dot{M}c^2}
    \sim 10^4\,\epsilon_{\rm e}\,\frac{\zeta_{\rm w}}{\zeta_{\rm d}}.
\end{equation}

The heated plasma quickly loses its energy via synchrotron cooling. 
The cooling timescale for particles with a Lorentz factor $\gamma\gg 1$ is given by 
\begin{equation}\label{eq:tcool}
   \tc\approx \frac{6\pi m_{\rm e} c}{\sigma_{\rm T} B^2\gamma}\approx \frac{8\times 10^{-5}\,{\rm s}}{\gamma_3\,B^2_5}.
\end{equation}
Here we normalized $\gamma$ to $10^3$ and the magnetic field in the disk $B$ to $10^5$~G. 
The magnetic field is expected to be comparable to the characteristic field in the pulsar wind at $R\sim 2\Rlc$,
\begin{equation}
  B_{\rm w}(R)\sim \frac{\mu}{R \Rlc^2}\approx 10^5 {\rm~G}.
\end{equation} 
The average wind pressure $\sim B_{\rm w}^2/8\pi$ is balanced by the disk pressure at the termination radius, and the disk magnetic pressure $B^2/8\pi$ is a fraction $\epsilon_B$ of the total pressure, so we rewrite $B$ in the form,
\begin{equation}
    B=\epsilon_B^{1/2}\,B_{\rm w}.
\end{equation}
The peak frequency of synchrotron emission by particles with Lorentz factor $\gamma$ is \citep{GS65}
\begin{equation}
\label{eq:nu}
    \nu_{\rm s}(\gamma)\approx 0.3\,\frac{3 e B}{4\pi m_{\rm e} c}\,\gamma^2\approx  10^{17}\,\gamma_3^2\,\epsilon_B^{1/2} {\rm ~Hz}.
\end{equation}
Most of the dissipated energy is radiated by particles with $\gamma\sim\gamma_{\rm m}$ and their emission falls in the X-ray band, explaining the high X-ray efficiency. 

In the simplest model, the heating mechanism produces a Maxwellian distribution with the thermal Lorentz factor $\gamma\sim\gamma_{\rm m}$. 
Such heating would occur in a magnetized shock that is quasi-parallel to the magnetic field \citep{Gallant92}.
As the particles are fast-cooling and their Lorentz factors $\gamma$ are decreasing below $\gamma_{\rm m}$, they radiate their energies $\gamma m_{\rm e}c^2$ at decreasing frequencies $\nu\sim\nu_{\rm s}(\gamma)$. 
The resulting spectral distribution of the emitted luminosity is given by
\begin{equation}
    \frac{dL}{d\ln\nu}\sim \zeta_{\rm d}\dot{N}\gamma m_{\rm e}c^2, \qquad \nu<\nu_{\rm m}\equiv\nu_{\rm s}(\gamma_{\rm m}),
\end{equation}
where $\gamma$ is related to $\nu\sim\nu_{\rm s}$ by Equation~(\ref{eq:nu}). 
This yields a simple estimate
\begin{equation}
\label{eq:sp}
   \frac{dL}{d\ln\nu}\sim \frac{\zeta_{\rm d} \,\dot{M} c^2}{\epsilon_B^{1/4}}\,\nu_{17}^{1/2}. 
\end{equation}
Substituting the estimated accretion rate of $\dot{M} \sim 10^{13} - 10^{14}${~g~s$^{-1}$}, we find the X-ray luminosity ($\nu\sim 10^{17}-10^{18}$~Hz) to be consistent with the observed $L_{\rm X}\sim 3\times 10^{33}$~erg~s$^{-1}$.

The X-ray spectrum predicted by Equation~(\ref{eq:sp}) has the standard fast-cooling photon index $\Gamma=1.5$. 
The observed spectrum is somewhat steeper, $\Gamma\approx 1.7$. 
This may be a problem for the simplest synchrotron model. 

If the heating mechanism produces a power-law particle spectrum with an index $p=-d\ln N/d\ln \gamma$, then fast cooling gives a synchrotron spectrum with photon index $\Gamma=1+p/2$.
The observed $\Gamma=1.7$ would be reproduced if $p=1.4$. 
However, it is unclear what dissipation mechanism gives an electron distribution with $p=1.4$, as typical slopes obtained for particle acceleration via shocks and magnetic reconnection in simulations are softer, $p\gtrsim2$ \citep[e.g.,][]{Spitkovsky08,SS14}.

In addition to the synchrotron cooling, the particles are cooled by the inverse Compton (IC) process. 
The ratio of the two cooling rates is
\begin{equation}\label{eq:IC}
  \frac{\dot{\gamma}_{\rm IC}}{\dot{\gamma}_{\rm s}}\approx \frac{U_{\rm rad}}{U_B}\sim \frac{0.1}{\epsilon_B},
\end{equation}
where $U_B=B^2/8\pi\sim 4\times 10^8\, \epsilon_B$~erg~cm$^{-3}$ is the magnetic energy density, and $U_{\rm rad}\sim L/4\pi R^2 c\sim 10^7 L_{33}$~erg~cm$^{-3}$ is the radiation energy density (in photons of sufficiently low energies, which scatter with Thomson cross section).
Hence, the electron losses to IC radiation are expected to be smaller, at least at the inner edge of the accretion disk where the magnetic field is strongest.
  
\begin{figure*}
\plotone{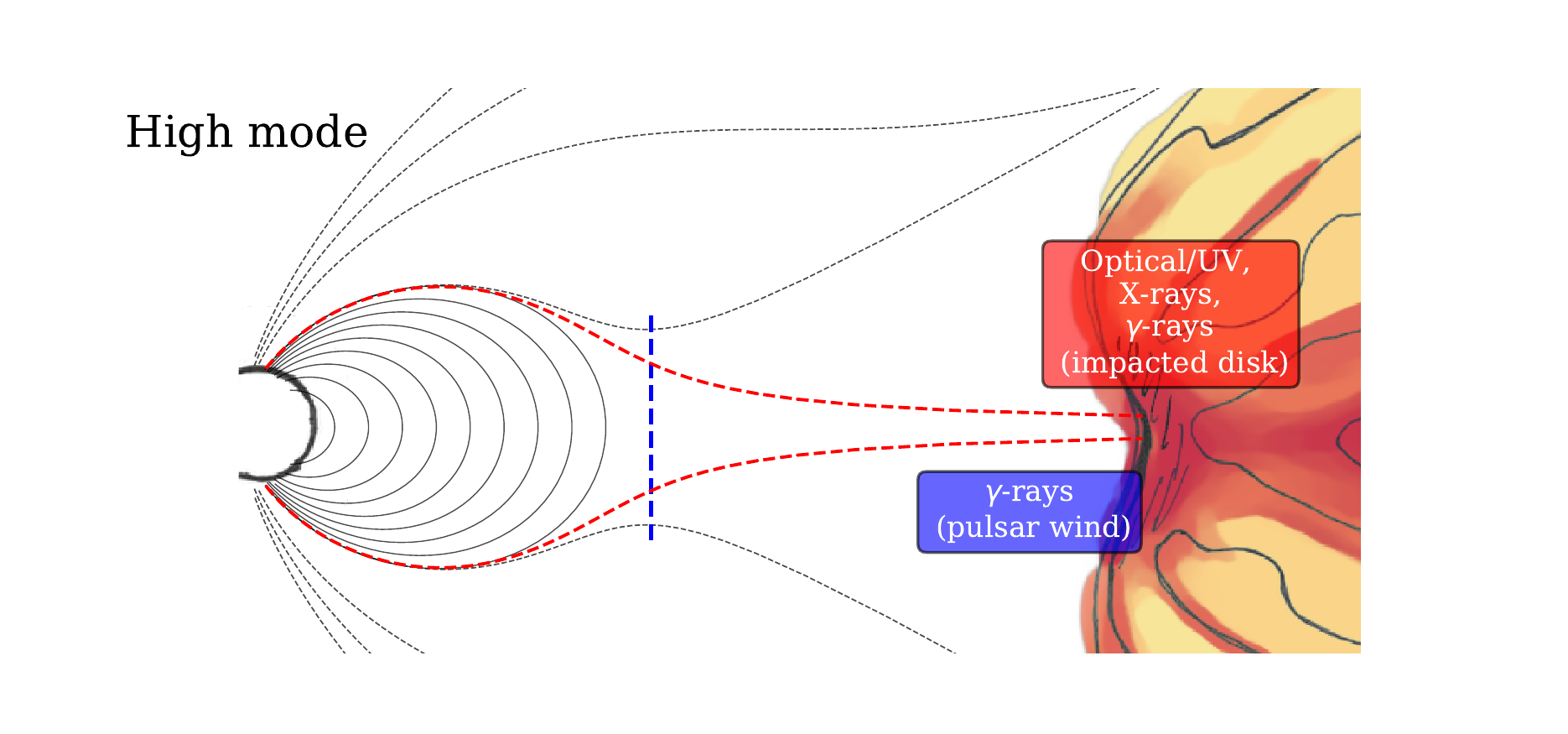}
\plotone{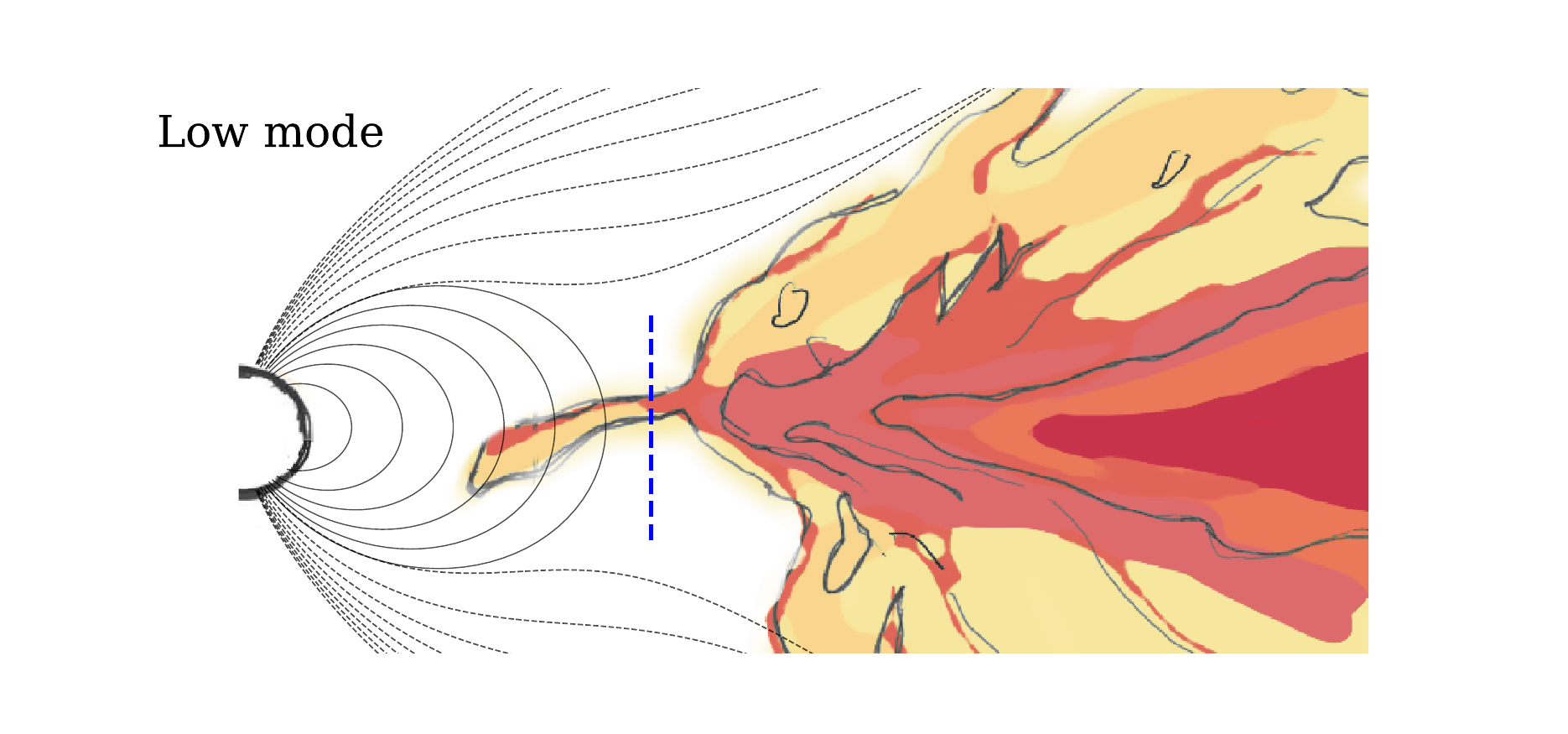}
\caption{
Schematic view of the high and low modes of PSR~J1023+0038.
Blue dashed lines show the location of the light cylinder. 
The pulsar magnetic field is shown by black curves (dotted for open field lines and solid for closed field lines).
Red and blue text boxes indicate emission components coming from the interaction of the pulsar wind with the truncated accretion disk
(colored in yellow-brown-red). \label{fig:rad_highlow}
}
\end{figure*}

\subsection{X-ray pulsations}\label{sect:xraypuls}

The observed X-ray pulsations at the pulsar spin frequency in our scenario are caused by a finite misalignment angle $\mis$ between the pulsar $\boldsymbol{\mu}$ and $\boldsymbol{\Omega}$. 
The rotational modulation is a subtle effect. 
We note that the total dissipated and emitted power in the wind-disk interaction remains constant as the inclined wind peak slides along the inner edge of the disk. 
Furthermore, since the emitting region is optically thin, the produced X-rays escape unaffected by the disk, in almost all directions except near the equatorial plane where the outer opaque disk can block the X-ray source
(see Figure~\ref{fig:rad_highlow}).
The reported disk inclination in PSR~J1023+0038, $i\approx 42\degr$, implies that the observer should see the synchrotron source all the time, regardless of the rotational phase.

Nevertheless, the observed luminosity should oscillate because of the anisotropic angular distribution of the produced synchrotron radiation.
The effect is particularly strong at large misalignment angles, $\tan \mis>H/R$, when the disk intercepts only a fraction of the current sheet.
Then there are two strong dissipation peaks confined to opposite regions of azimuthal angle $\phi$, and this pattern rotates with the pulsar period.
Let $\phi_{\rm p}=\Omega t$ and $\phi_{\rm p}+\pi$ be the azimuthal positions of the two dissipation peaks of the wind-disk collision. 
We note that the vector of the disk magnetic field $\boldsymbol{B}$ is not constant with $\phi$, even in an axisymmetric disk. 
As an example, consider an idealized disk with a purely toroidal magnetic field. 
Then the direction of $\boldsymbol{B}$ turns with $\phi$ through $2\pi$ angle around the disk axis, and angle $\psi$ between $\boldsymbol{B}$ and the observer's line of sight changes in the range 
\begin{equation}
    \frac{\pi}{2}-i<\psi< \frac{\pi}{2}+i.
\end{equation}
Isotropically heated electrons emit anisotropic synchrotron radiation. 
It is maximum in the direction perpendicular to the local $\boldsymbol{B}$ and vanishes along $\boldsymbol{B}$. 
The observed luminosity is proportional to the angular factor $\sin^2\psi$, and in our example this factor varies between unity and $\cos^2i$.
In an idealized picture where the wind-disk collision is localized at $\phi_{\rm p}=\Omega t$ and $\phi_{\rm p}+\pi$, and vanishes at all other $\phi$, the observed synchrotron emission oscillates with the pulsed fraction $(1-\cos^2i)/(1+\cos^2 i)$, reaching maximum when the rotating $\boldsymbol{B}$ is perpendicular to the line of sight.

In a more realistic model, the oscillation amplitude is somewhat reduced, because the wind-disk collision occurs at all $\phi$ outside $\phi_{\rm p}$, although with a reduced power. 
Furthermore, the disk magnetic field can have a significant random component. 
In a tangled small-scale field, the synchrotron emissivity would become isotropic and pulsations would be suppressed. 
Thus, the presence of a large-scale magnetic field in the heated region is essential for producing X-ray pulsations in our model. 
Such a large-scale field can indeed be generated in accretion disk driven by magnetorotational instability \citep[e.g.,][]{ONeill2011}. 
We conclude that it is reasonable to expect a pulsation amplitude of $\sim$10\%. 
The observed root-mean-square amplitude of the X-ray pulsations in PSR~J1023+0038 is $\sim$6\%.

The pulse profile predicted by our model should have two peaks separated by half of the rotation period, as observed in PSR~J1023+0038. 
If the disk is axisymmetric and the pulsar has a centered magnetic dipole moment $\boldsymbol{\mu}$, the peaks should have equal amplitudes. 
A more complicated pulsar magnetosphere becomes approximately dipole at distances $R\gg R_{\rm NS}$, with deviations $\sim R_{\rm NS}/R$. 
Thus, at radii $R\sim 10R_{\rm NS}\gtrsim \Rlc$, one can expect deviations of $\sim 10\%$ from the idealized picture of a rotating dipole.  
They can lead to $\sim 10$\% asymmetry of the two peaks of wind-disk collision at $\phi_{\rm p}$ and $\phi_{\rm p}+\pi$. 
It will result in somewhat unequal amplitudes of the two peaks in the pulse profile.

\subsection{Optical pulsations}

The spectrum of synchrotron emission extends from the X-ray band to lower frequencies, including the optical band, in particular $\nu\approx 6.5\times 10^{14}$~Hz, at which PSR~J1023+0038 was observed \citep{Ambrosino17}.
Two additional effects should be examined for the low-frequency emission. 

First, the emission below $10^{15}$~Hz is produced by electrons with a lower Lorentz factor $\gamma<10^2$, which have a longer cooling timescale $\tc$, see Equations~(\ref{eq:tcool}) and (\ref{eq:nu}). 
This timescale should be compared to the period of pulsar rotation that modulates the observed synchrotron flux and to the residence time of the heated plasma in the emission region. 
The strong heating of the disk plasma by the pulsar wind results in its ejection. 
The ejection speed may be roughly estimated as the sound speed, which is controlled by the ion temperature $T_{\rm i}$. 
Ions receive the fraction $(1-\epsilon_e)$ of the dissipated power, and $T_{\rm i}$ immediately after heating is given by 
\begin{equation}
   kT_{\rm i}\sim (1-\epsilon_{\rm e}) \,\frac{\zeta_{\rm w}\Lsd m_{\rm p}}{\zeta_{\rm d} \dot{M}}.
\end{equation}
It gives a moderately relativistic $kT_{\rm i}\sim m_{\rm p}c^2$ for the typical parameters estimated in Sections~\ref{sect:ener} and \ref{sect:emis}.
The ions are unable to radiate their energy or pass it to the electrons via Coulomb collisions on timescales of interest, because of the low plasma density. 
Therefore, the sound speed in the heated disk plasma likely remains comparable to $c_{\rm s}\sim 10^{10}$~cm~s$^{-1}$ even after the electrons radiate most of the received heat. 
We roughly estimate the ejection/residence timescale as 
\begin{equation}
  t_{\rm esc}\sim \frac{R}{c_{\rm s}}\sim 1 {\rm~ms}.
\end{equation}
Incidentally, this timescale is comparable to the pulsar rotation period.

The fast-cooling assumption is approximately valid for electrons with $\tc<1$~ms, which corresponds to Lorentz factors $\gamma_{\rm c}>80\,B_5^2$ and $\nu>\nu_{\rm c}\equiv\nu_{\rm s}(\gamma_{\rm c})\sim 8\times 10^{14}B_5^{-3}$~Hz. 
The fact that $\nu_{\rm c}$ is close to the optical band implies that the optical flux from the wind-disk collision should not be far from the simple extrapolation of the fast-cooling X-ray spectrum to the optical band.
If the photon spectral index remains at $\Gamma\approx 1.7$ over the extended range of frequencies, the optical luminosity should be $L_{\rm opt}\sim 0.1L_{\rm X}\sim 3\times 10^{32}$~erg~s$^{-1}$. 
The observed total optical emission is brighter because of the contribution of additional components (e.g., coming from the reprocessed X-ray emission), however, the {\it pulsed} emission is approximately consistent with a single power law observed in the X-ray band and extended down to the optical band \citep[e.g.,][]{Papitto19}.
At frequencies $\nu\ll 10^{15}$~Hz the spectral slope is expected to change --- the emission should be reduced due to inefficient cooling, as the electron energy $\sim \gamma_{\rm c} m_{\rm e}c^2$ is lost to adiabatic cooling of the ejected plasma rather than radiated.

At even lower frequencies, the synchrotron emission is affected by the self-absorption. 
The brightness temperature of emission with luminosity $dL/d\ln\nu$ is
\begin{equation}
   kT_{\rm b}(\nu)\sim \frac{c^2}{8\pi^2 R^2 \nu^3}\,\frac{dL}{d\ln\nu}. 
\end{equation}
Substituting $\nu\sim 6\times 10^{14}$~Hz and our estimate $dL/d\ln\nu\sim 3\times 10^{32}$~erg~s$^{-1}$, we find $kT_{\rm b}/m_{\rm e}c^2\sim 2\times 10^{-2}$, much smaller than the Lorentz factor of the electrons emitting the optical radiation, $\gamma\sim80$.
Therefore, self-absorption is negligible in the optical band in our model.
It becomes important in the infrared band, at frequencies $\nu\lesssim10^{14}$~Hz.

The optical synchrotron emission should have a pulse profile with two peaks separated by half of the rotation period, similar to the X-ray pulse profile. 
The amplitudes of the two peaks can differ by $\sim 10$\%, if the pulsar magnetosphere is more complicated than a centered dipole (see the end of Section~\ref{sect:xraypuls}). 
Such unequal peaks separated by half of the period have been observed in PSR~J1023+0038 \citep{Papitto19}.

In the model of impulsive electron heating to $\gamma_m\sim 10^3-10^4$, 
the electrons cool down to $\gamma\sim 10^2$, at which synchrotron emission peaks in the optical band, on the timescale $\lesssim$1~ms.
Then the optical emission will peak with a delay, which is controlled by radiative and adiabatic cooling of the plasma expanding away from the heating site.
Thus, the optical pulses can be shifted relative to the X-ray pulses by $\lesssim$1~ms.
A small phase difference between the optical and X-ray pulsations has been reported in PSR~J1023+0038 \citep{Papitto19}.

\section{Discussion}\label{sect:discus}

\subsection{Flaring mode}

X-ray flares can be caused by rapid, strong increase of the disk thickness at the inner edge (Figure~\ref{fig:geom_flare}).
This allows the disk to intercept (and radiate) a larger fraction $\zeta_{\rm w}$ of the pulsar wind.
When the disk thickness reaches $H/R\sim 0.5$, it begins to intercept most of the wind power. 
For instance, a weakly misaligned rotator emits most of the spindown power (Poynting flux) at $\sim30\degr$ from the equatorial plane \citep{ChenBeloborodov14}, and thus it will be intercepted by the inflated disk.
Dissipation of about half of the spindown power of PSR~J1023+0038, $L_{\rm sd}\approx 4.4\times 10^{34}$~erg~s$^{-1}$, is sufficient to explain the observed luminosity of the strongest X-ray flares in the 3--70~keV band, $L_{\rm X}\approx  1.2\times10^{34}$~erg~s$^{-1}$.
We note that this luminosity is a factor of 100 higher than in the RP state \citep{Archibald10,Bogdanov11,Tendulkar14,Bogdanov15}, demonstrating that the wind-disk collision is far more efficient in converting $L_{\rm sd}$ to X-rays compared to the isolated pulsar.

The increased disk thickness also explains the observed changes in pulsation behavior. 
In the flaring mode, the X-ray pulsations are barely detected at the $2\sigma$ level \citep{Archibald09,Bogdanov15}. 
Optical pulsations are clearly present  ($\sim8\sigma$ detection, see \citealt{Papitto19}) with a pulse profile similar to that in the high mode, however, the pulsed luminosity is reduced by a factor of $\sim 5$. 
The strong reduction of pulsations is expected in the outlined scenario for the flares: when the disk becomes thick, the azimuthal variation of the intercepted wind power decreases, which reduces the pulsation amplitude of synchrotron emission without changing much its double-peak shape.

The physical mechanism  for sudden inflation of the disk inner edge still needs to be understood.
The disk may inflate as matter piles up at its inner edge and becomes squashed by the wind pressure at the inner edge (facing the neutron star). 
The increased effective height of the disk might be caused by rearrangement of the magnetic structure within the flow, e.g. in the process of the magnetic loop inflation \citep{RUK98}. 
The loops present an obstacle to the pulsar wind, and their interaction may be dissipative. 
Another possibility for the increased disk thickness is an unstable thermal balance.

\subsection{Duration of modes and timescales of transitions}

PSR~J1023+0038 typically spends $\sim$100--600~s in the high mode and $\lesssim$50~s in the low mode. 
The time separation between the low-mode episodes follows a log-normal distribution \citep{Tendulkar14,Bogdanov15}, similar to accretion rate fluctuations inferred for black hole accretion disks \citep{Gaskell04,UttleyMV05}.
The transitions may be interpreted as a hysteresis-type response to variations in $\dot M$: 
the disk penetrates the light cylinder when the mass accretion rate exceeds a threshold $\dot M_{\rm c1}$, 
and the reverse transition occurs when the accretion rate $\dot M$ drops below another threshold $\dot M_{\rm c2}<\dot M_{c1}$.
Then the time the pulsar spends in each mode is
determined by variations in $\dot M$ and so inherits its statistical properties, such as the log-normal distribution.

A natural timescale for the transition from the high mode to the low mode is the viscous timescale at the disk inner edge, $R\sim 2R_{\rm lc}$.
This timescale is $t_{\rm visc}=\alpha^{-1}(H/R)^{-2} t_{\rm K}\sim 10\,(R/10H)^2$~s, where $t_{\rm K}=(R^3/GM_{\rm NS})^{1/2}\sim10^{-2}$~s is the Keplerian rotation timescale, and we assumed a typical viscosity parameter $\alpha\sim0.1$.
The observed transitions between the high and low modes of PSR~J1023+0038 occur in $\sim$10--20~s, in both X-ray and optical bands \citep{Bogdanov15,Shahbaz15}. 
This is consistent with $t_{\rm visc}$.
The high-to-low mode transition is observed to occur faster, on average, than the reverse transition.
Such an asymmetry is consistent with our scenario, as the disk evacuation from the light cylinder (low-to-high mode transition) may proceed slower than the penetration (high-to-low mode transition).

The observed fractional amplitude of variations of the X-ray luminosity on short, seconds to minutes, timescales is higher \citep{Tendulkar14}.
This indicates a lower stability of the ``propeller'' wind-disk interaction inside the light cylinder, consistent with variability observed in numerical simulations of \citet{PT17}.  
There is also variability on much longer timescales, from one observing run to another.
For the high mode, the average $L_{\rm X}$ varies between the runs in the modest range of $L_{\rm X}=(3.0-3.3)\times10^{33}$~erg~s$^{-1}$, and the low mode exhibits stronger variations, $L_{\rm X}=(3.4-5.4)\times 10^{32}$~erg~s$^{-1}$ \citep[table~1 in][]{Jaodand16}. 

The low mode variations are comparable with the typical 20--25\% variations in accretion disks of black hole X-ray binaries \citep[e.g.,][]{UM05}; they are not surprising.
On the other hand, the remarkable stability of the high-mode $L_{\rm X}$ remains puzzling. 
In our scenario, it requires the disk thickness in the high mode to stay within $\sim 10$\% at a preferred value $H/R\sim0.1$. 
Since $H/R$ is shaped by the wind colliding with and flowing around the disk, a preferred balanced value $H/R\sim 0.1$ seems plausible. 
However, the observed flares require occasional catastrophic loss of this balance, so that $H/R$ jumps by a factor of $\sim4-6$. 
The flares are detected during the high X-ray mode, although they are sometimes preceded by a low-mode interval \citep{Bogdanov15}.
The shortest flares last less than a minute; they sometimes come in sequence with the total duration of the order of 45~min, both in optical and X-rays. 
The ingress/egress timescales of flares are comparable to those of mode switching, $\lesssim$30~s.

\subsection{Gamma-ray emission}

The observed average $\gamma$-ray flux increased five-fold as the pulsar transited to the LMXB state \citep{Stappers14}. 
The $\gamma$-ray luminosity is comparable to the high-mode X-ray luminosity. 
Its spectrum has a break above $\sim1$~GeV \citep{Takata14,Papitto19}.

Our picture for the high mode predicts two bright $\gamma$-ray sources in the region of pulsar wind collision with the disk (see Figure~\ref{fig:rad_highlow}):
\\
(1) The pulsar wind heated by its collision with the disk emits high-energy synchrotron photons. 
The Lorentz factors of the heated wind particles can reach $\gamma\sim 10^6-10^7$, depending on the $e^\pm$ multiplicity of the wind (Equation~\ref{eq:gamma_mw}). 
Their synchrotron emission peaks at $h\nu_{\rm s}\gtrsim 1$~GeV and their luminosity is a large fraction of the dissipated power $\zeta_{\rm w}\Lsd$, consistent with the observed gamma-ray emission.
\\
(2) The disk plasma heated by the pulsar wind. 
Besides emitting the synchrotron X-rays, this plasma produces an IC component which falls into the gamma-ray band. 
The high-energy break of the IC spectrum is shaped by Klein-Nishina effects and reflects the $\gamma_{\rm m}$ peak of the electron distribution: $h\nu\sim\gamma_m m_{\rm e} c^2/2$. 
It is related to the synchrotron spectrum break, which may be at $\sim50$~keV \citep[from the {\it NuSTAR} data, see][]{Tendulkar14,Papitto19}, and corresponds to $\gamma_{\rm m}\sim10^4$ in our model. 
This gives the gamma-ray break at $2-3$~GeV. 
The ratio of the IC and synchrotron luminosities was estimated in Equation~(\ref{eq:IC}) as $0.1/\epsilon_B$. 
Hence, the IC $\gamma$-ray luminosity can be consistent with observations if $\epsilon_B\sim 0.1$.

\subsection{UV, optical and infrared emission}

The pulsed optical emission from PSR~J1023+0038 is consistent with the extrapolation of its pulsed X-ray spectrum. 
However, the extrapolation does not hold for the unpulsed component. 
The total optical luminosity is an order of magnitude above the extrapolation of the X-ray power-law to the optical band, in both high and low mode \citep{Bogdanov15,Papitto19}. 
This is expected: a big fraction of optical luminosity comes from the reprocessing of the X-rays by the outer regions of the accretion disk and the companion star, as evidenced by the large modulation with the binary orbital period \citep{Shahbaz15}.

A simple reprocessing model would predict that the optical luminosity should vary in direct response to the X-ray luminosity, with a delay on the light-crossing timescale of the system. 
This picture is supported by simultaneous optical and X-ray observations taken in May 2017 \citep{Papitto19}. 
The cross-correlation function (CCF) measured in this period showed a strong positive correlation between the optical and X-ray flux variations.
During the high and low modes, the CCF had a single and broad (width $\sim$5~s) positive peak around a few second delay, consistent with reprocessing on scales comparable to the binary separation of $\sim$1 light second.
In addition, in the high mode, there is a hint of a correlation at shorter timescales $<1$~s, which may be associated with the synchrotron component generated by the pulsar wind-disk collision.

PSR~J1023+0038 occasionally emits optical flares, which were found to correlate with the X-ray flares \citep{Bogdanov15,Jaodand16,Papitto19}.
The optical/X-ray CCF of the flaring mode was found to be narrower than the CCF in the high/low X-ray modes, and its peak was consistent with zero lag (the lag is constrained to be smaller than 1~s). 
This change can result from the increased weight of the synchrotron optical radiation in the observed variability.\footnote{The synchrotron optical emission of the flare scales linearly with $L_{\rm X}$ (assuming a fixed spectral slope), while the reprocessed optical emission has a weaker dependence on $L_{\rm X}$.
}

However, numerous existing observations of the system showed a more complicated picture.
While some datasets indicate a positive correlation between the X-ray and UV/optical flux variations \citep{CotiZelati18,Papitto19}, 
others show an anti-correlation \citep[figure~7 of][]{Jaodand16}.
Furthermore, observations taken in February 2017 showed that the system spent most of the time at a lower optical flux \citep[see figure~2 in][no simultaneous X-ray data is available]{Shahbaz18}.
This is opposite to the usual X-ray behavior and could be interpreted as the low optical mode appearing during the high X-ray mode.
Several factors may contribute to this behavior, including X-ray anisotropy and the presence of another optical/infrared component.
The extra low-frequency component may be emitted by the hot accretion disk outside the truncation radius \citep{VPV13}, before its interaction with the pulsar wind.
Contribution of the hot disk emission to the total spectrum grows with decreasing frequency and further complicates the observed variability.

The long-term monitoring of PSR~J1023+0038 with \textit{Kepler} mission (effective wavelength passband 420-900~nm) showed that the low/high mode variability pattern changed with time \citep{Kennedy18}.
Some segments of the \textit{Kepler} light curve exhibit a bimodal flux distribution while others show one preferential flux with some flickering around it.
A similar complexity is found in the infrared band. 
Two runs of infrared observations in February and May 2017 showed two different distributions of fluxes --- single-peak and bimodal \citep{Shahbaz18,Papitto19}. 
This complexity can be explained by the time-dependent contribution of the hot disk to the observed flux. 
It could also explain the so-called precognition dip observed in the infrared/optical CCF \citep{Shahbaz18}.

\subsection{Radio emission}

Simultaneous radio/X-ray observations showed enhancement of radio emission every time the system transited to the low mode \citep{Bogdanov18}.
The accretion picture described in Section~\ref{sect:pheno}
offers a natural explanation for this enhancement.
The disk penetration into the light cylinder, which we associate with the transition to the low mode, should increase the open magnetic flux of the pulsar and hence increase its wind power \citep{Parfrey16,PT17}. 
We argued that the wind-disk interaction is less dissipative in this configuration, because the disk occupies the Y-point, so that the wind flows around the disk rather than collides with it. 
Most of the spindown power flows out from the system. 

The change of the disk-wind interaction from ``collision'' (high mode) to ``propeller'' (low mode) is also likely accompanied by an enhancement in the matter loading of the relativistic outflow. 
As the outflowing material expands, it becomes more rarefied and transparent to synchrotron radiation at low frequencies leading to an increase of the infrared and radio luminosity, as observed in the low mode.

\section{Summary}\label{sect:summary}
 
In this paper we discussed radiative signatures of accretion at a low rate onto a millisecond pulsar. 
We argued that when the pulsar wind truncates the accretion disk outside the light cylinder, the system is capable of efficiently producing X-rays. 
The X-ray luminosity is sustained by the pulsar spindown power rather than the gravitational power released by accretion.
The disk edge outside the light cylinder $\Rlc$ forms an obstacle for the relativistic pulsar wind and is heated by the collision with it. 
It intercepts, in particular, the fan-like peak of the wind power, which wobbles around the equatorial plane as the pulsar rotates. 

In contrast, when the disk penetrates the light cylinder and occupies the magnetospheric Y-point, it shuts down the equatorial fan of the pulsar power.
Then the disk interacts mainly with the closed magnetosphere. 
Their interaction results from the mismatch of rotational velocities and is of the ``propeller'' type.
Matter ejection by the propeller inside $\Rlc$ can be significantly less dissipative than the wind-disk collision outside $\Rlc$.

The two accretion regimes, with the collision and propeller types of pulsar-disk interaction, produce X-rays with different efficiencies. 
This may explain the two X-ray modes exhibited by PSR~J1023+0038.
We suggest that in both modes the disk accretes with a modest rate $\dot{M}$ between $10^{13}$ and $10^{14}$~g~s$^{-1}$, and therefore is truncated near the light cylinder, far outside the co-rotation radius. 
The high mode is powered by the wind-disk collision outside $\Rlc$. 
The low-mode episodes occur when the disk occasionally penetrates the light cylinder and shuts down the equatorial wind.
The low-mode luminosity can be sustained by both accretion and the pulsar spindown power.
Transitions between the low and high modes do not require big changes in $\dot{M}$ and may result from strong variability in the pulsar-disk interaction, which is seen in numerical simulations \citep{PT17}.

The proposed picture is consistent with a number of observed properties of PSR~J1023+0038:

\begin{itemize}
\item The accretion disk weakly affects the pulsar spindown rate, because it stays outside $\Rlc$ most of the time. 
This explains why the spindown of PSR~J1023+0038 in the accretion state (averaged over the high and low modes) is only $\sim$27\% stronger than in the radio-pulsar state.
\item Disk particles are heated in the wind-disk collision to Lorentz factors $\gamma\sim 10^3-10^4$ and produce synchrotron emission.
It  peaks in the X-ray band, explaining the high X-ray efficiency of the system in the high mode. 
\item The X-ray emission occurs in the fast-cooling regime and tracks the wind modulation by the pulsar rotation.
This modulation produces a pulse profile with two peaks separated by 180\degr, consistent with the observed pulse profile.   
\item The synchrotron spectrum extends to lower frequencies, explaining the pulsed component of the optical/UV luminosity, with the optical pulse profile similar to the X-ray one.
\item The wind-heated disk also produces inverse Compton emission in the GeV band.
Furthermore, synchrotron GeV $\gamma$-rays are expected from the $e^\pm$ particles of the wind energized by its collision with the disk. 
The expected gamma-ray luminosity and the GeV spectral cutoff are consistent with {\it Fermi} observations of PSR~J1023+0038.
\item X-ray flares may result from sudden/catastrophic increases of the effective scale-height of the accretion disk intercepting the pulsar wind. 
Such events are expected to boost the synchrotron emission and reduce pulsations, as observed. 
\item When the disk penetrates the light cylinder, it cuts open part of the closed magnetosphere. 
This may explain the higher radio activity in the low mode.
\item 
The observed complex variability of the system at the optical and infrared wavelengths can be caused by the presence of a few different components, in particular the reprocessed radiation and the internal emission from the hot disk outside the truncation radius.
\end{itemize}

The presented estimates support the outlined accretion picture for PSR~J1023+0038.
However, they do not provide a detailed description for the disk heating by the pulsar wind.
A simple model of shock heating followed by fast synchrotron cooling of the Maxwellian postshock plasma would predict a photon index $\Gamma=1.5$, inconsistent with the observed $\Gamma\approx 1.7$.
Dedicated numerical simulations can help understand the particle distribution function in the dissipation region.

Future theoretical work will be helped by anticipated advanced observations. 
New X-ray detectors such as \textit{STROBE-X} \citep{Ray19} and \textit{eXTP} \citep{Zhang2018} will have large collecting areas and provide much better photon statistics for dim objects like PSR~J1023+0038. 
This will give accurate measurements of the pulse profile and its spectral variation.

\bigskip

\section*{Acknowledgements}

The authors thank Juri Poutanen, Kyle Parfrey, Jari Kajava, and Slavko Bogdanov for useful discussions.
AV acknowledges the Academy of Finland grant 309308.
The work was supported by the grant 14.W03.31.0021 of the Ministry of Science and Higher Education of the Russian Federation.
AMB is supported by NASA grant NNX17AK37G, a Simons Investigator Award (grant \#446228), and the Humboldt Foundation.

\bibliographystyle{aasjournal.bst}



\end{document}